# Dynamical decoupling of a geometric qubit

Yuhei Sekiguchi, Yusuke Komura, and Hideo Kosaka*

Yokohama National University, 79-5 Tokiwadai, Hodogaya, Yokohama 240-8501, Japan

*kosaka-hideo-yp@ynu.ac.jp

## Abstract

**Quantum bits or qubits naturally decohere by becoming entangled with uncontrollable environments. Dynamical decoupling is thereby required to disentangle qubits from an environment by periodically reversing the qubit bases, but this causes rotation error to accumulate. Whereas a conventional qubit is rotated within the SU(2) two-level system, a geometric qubit defined in the degenerate subspace of a V-shaped SU(3) three-level system is geometrically rotated via the third ancillary level to acquire a geometric phase. We here demonstrate that, simply by introducing detuning, the dynamical decoupling of the geometric qubit on a spin triplet electron in a nitrogen-vacancy center in diamond can be made to spontaneously suppress error accumulation. The geometric dynamical decoupling extends the coherence time of the geometric qubit up to 1.9 ms, limited by the relaxation time, with 128 decoupling gates at room temperature. Our technique opens a route to holonomic quantum memory for use in various quantum applications requiring sequential operations.**

# Main text

Quantum information technology is becoming a reality in the form of quantum computers, simulators, sensors, as well as the repeaters required for the quantum internet. Long memory time and high-fidelity gates are the key factors that must be scaled up to finally achieve real applications. The widely used dynamical decoupling technique [1–8], which in principle extends the memory time or the coherence time, in practice faces the problem of error accumulation after a large number of decoupling gates, which eventually degrades the state fidelity. The Carr-Purcell-Meiboom-Gill (CPMG) sequence [1] has thus been developed to suppress the accumulation of gate errors, while the initial state is restricted to the eigenstate of the driving field of the decoupling gate. As alternatives, a composite pulse technique for achieving high-fidelity gates [3,4] and a specially designed gate sequence [2,5,6] have been developed to be independent of the initial state.

A qubit is typically defined as being in a two-level system with an energy gap, which allows direct transition within the bases to implement dynamic quantum gates. Another type of qubit can also be defined in a two-level system without an energy gap; this type requires an indirect transition via a third ancillary level, and thus constitutes a V-shaped three-level system to implement geometric quantum gates [9–12]. Geometric quantum gates can be either adiabatic [13–15] or non-adiabatic [9–12,16–19]. In contrast to the adiabatic geometric gate, the non-adiabatic geometric gate enables faster gate operation to reduce the influence of the environmental noise, thereby resulting in high fidelity. Moreover, the degenerate two-level system is independent of the global phase of the driving field, as seen in the polarization or time-bin encoding of a photon, enabling post selection of successful operations to exclude a population loss from the qubit space spanned by the degenerate two-level system.

The negatively charged nitrogen-vacancy (NV) center in diamond (Fig. 1a) naturally consists of a V-shaped spin-1 electron system (Fig. 1b), which is a promising platform for quantum sensing of electromagnetic fields and temperatures to probe nanoscale physical, chemical, and biological phenomena [20,21] and for the quantum repeaters interfacing between a photon and a solid-state spin for long-distance quantum communication [22–25]. In contrast to conventional SU(2) gates, which utilize the global phase of the driving field to achieve arbitrary rotation, SU(3) gates utilize the polarization of the driving field for the demonstration of spin-photon entanglement generation [22,23] and quantum state transfer [24]. Polarization-based geometric spin rotation [10–12] has also been demonstrated under a zero magnetic field, where the electron spin coherence time is maximized owing to the time reversibility of the system including environmental spins [9]. However, it is likely that the population leaks from the qubit space to the ancillary space during the geometric gate operations as an additional error [26,27], limiting the previous experiments to several gate operations. Here we demonstrate robust dynamical decoupling of a geometric qubit, which decouples the qubit state not only from environmental spins but also from the ancillary state, with a spin-1 electron spin system in an NV center in diamond.

The spin-1 electron spin system of the NV center consists of degenerate $|m_s = \pm 1\rangle$ levels and a $|m_s = 0\rangle$ level, which are mutually split in energy by the zero-field splitting. Instead of the conventional dynamic qubit based on the $|0\rangle$ state and one of the $|\pm 1\rangle$ states, which are energetically split under a magnetic field, we define the geometric qubit based on the degenerate $|\pm 1\rangle$ states serving as an interface between an electron and photon polarization. The Hamiltonian of the geometric qubit with the x-polarized microwave under a zero static field is expressed by $H_{\text{drive}} = \frac{\Omega}{2} S_x$, where $\Omega$ denotes the Rabi frequency and $S_x$ denotes the x-component of the spin-1 operator. The Hamiltonian defines the bright state $|+\rangle = (|+1\rangle + |-1\rangle)/\sqrt{2}$, which is coupled to the ancilla state $|0\rangle$, and the dark state $|-\rangle = (|+1\rangle - |-1\rangle)/\sqrt{2}$, which is the uncoupled eigenstate of the Hamiltonian. Note that we hereafter omit the Dirac constant $\hbar$ for simplicity. A cyclic rotation in the operational space based on $\{|0\rangle, |+\rangle\}$ induces a geometric phase $\pi$ as a global phase on the bright state $|+\rangle$ to implement the geometric bit-flip gate in the qubit space $\{|+1\rangle, |-1\rangle\}$, leading to the dynamical decoupling of the geometric qubit. In practice, the pulse length error causes deviation of the rotation angle (angle error) to result in direct population leakage from the qubit space or dynamic rotation from $|+\rangle$ to $|0\rangle$. On the other hand, the state splitting between $|+1\rangle$ and $|-1\rangle$, including

the Zeeman splitting induced by a residual magnetic field and hyperfine splitting induced by the environmental spins, causes the deviation of the rotation axis (axis error) to result in indirect population leakage from the qubit space or geometric rotation from $|-\rangle$ to $|0\rangle$ via $|+\rangle$ (Fig. 1d) [28]. The angle error of the dynamic rotation accumulates every time the dynamical decoupling gate is operated (Fig. 1e left), while the axis error of geometric rotation is canceled out every two rotations [28].

The key feature of the demonstration is that the degeneracy between the qubit space and ancillary space under the microwave is lifted by simply introducing a frequency detuning $\Delta$, which is sufficiently smaller than $\Omega$. When the geometric bit-flip gates are periodically operated in the decoupling sequence (Fig. 1c), the angle error is averaged out by the phase acquired during the gate interval time $\tau$ (Fig. 1d) only when $\tau$ does not resonate to the detuning as $\tau \neq \frac{2\pi n}{\Delta}$ ($n \in \mathbb{N}$) (Fig. 1e middle). In contrast, the angle error accumulates when $\tau$ resonates to the detuning as $\tau = \frac{2\pi n}{\Delta}$ (Fig. 1e right), where we can estimate $\Delta$ from the $\tau$ dependence of the state fidelity. Note that this method is unique to the geometric qubit, since the qubit space and the operational space are separated, and the difference in the global phase of the microwave does not essentially change the rotation axis of the geometric qubit.

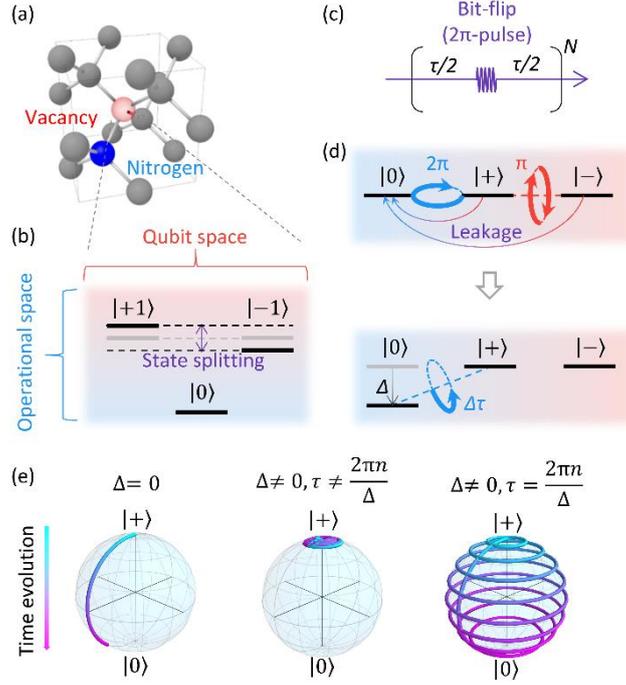

FIG. 1 (a) Schematic illustration of a single nitrogen-vacancy center in diamond. (b) V-shaped three-level system of a spin-1 electron spin. (c) Gate sequence of the dynamical decoupling of the geometric qubit. *N* indicates the number of decoupling gates. (d) Transformed level diagram on the rotating frame of x-polarized microwave. A $2\pi$-rotation in the operational space spanned by the bright state $|+\rangle = (|+1\rangle + |-1\rangle)/\sqrt{2}$ and the ancillary state $|0\rangle$ generates the geometric phase $\pi$ in the qubit space spanned by $|+\rangle$ and the dark state $|-\rangle = (|+1\rangle - |-1\rangle)/\sqrt{2}$. This rotation serves as a bit-flip gate, while population leakage from the qubit space to the ancillary space is induced by the angle error and axis error of the rotation (top). When a detuning $\Delta$ between the qubit space and the ancillary space is introduced, the relative phase shift between the qubit and ancillary spaces is accumulated during gate interval time $\tau$ (bottom). (e) Schematic time evolution of $|+\rangle$ in the operational space during the dynamical decoupling of the geometric qubit under a zero detuning (left), finite detuning without the resonant condition $\tau \neq \frac{2\pi n}{\Delta}$ ($n \in \mathbb{N}$) (middle), and finite detuning with the resonant condition $\tau = \frac{2\pi n}{\Delta}$ (right).

We use a single NV center consisting of a nitrogen $^{14}$N and a vacancy in type-IIa diamond with a natural abundance of $^{13}$C at room temperature. The experimental setup is the same as that in Ref. [9]. An external magnetic field was applied to carefully compensate for the geomagnetic field of ~0.045 mT using a permanent magnet with monitoring of the optically detected magnetic resonance (ODMR) spectrum within 0.1 MHz. The geometric bit-flip gate is performed with $\Omega = (2\pi) \times 25$ MHz, which is not sufficiently larger than the hyperfine splitting induced by the $^{14}$N nuclear spin of $(2\pi) \times 2.2$ MHz. Under the condition without detuning, the numerical calculation indicates that the population leakage of $|+\rangle$ after the dynamical decoupling of the geometric qubit increases with the number of gates due to the gate imperfection caused by the state splitting (Fig. 2a). On the other hand, when the detuning $\Delta = (2\pi) \times 130$ kHz is provided, the population leakage is suppressed except for $\tau = n \times 7.7$ μs, where the gate interval time $\tau$ resonates to the detuning (Fig. 2b). Figure 2c shows the experimental results for the dynamical decoupling of the geometric qubit, whose envelopes agree well with the numerical simulations, indicating a significant improvement in the state population under the non-resonant conditions. The absence of high-frequency components based on the $^{14}$N nuclear spin indicates an inhomogeneous broadening of the state splitting induced by the hyperfine interaction with the carbon $^{13}$C nuclear spin bath. The fluctuation of the dip position is caused by the temperature dependence of the zero field splitting 74 kHz/K [29], which shifts the detuning. The temperature fluctuation is around ± 0.2 K, which corresponds to ± 15 kHz. The dip width is narrower for larger $N$, indicating that the dip position is more sensitive to the gate interval time due to the longer exposure time under a detuning. On the condition that the accumulated angle error after $N$ decoupling gates is smaller than π, a narrower dip can be obtained by increasing $N$. The dip further narrows by increasing τ [28].

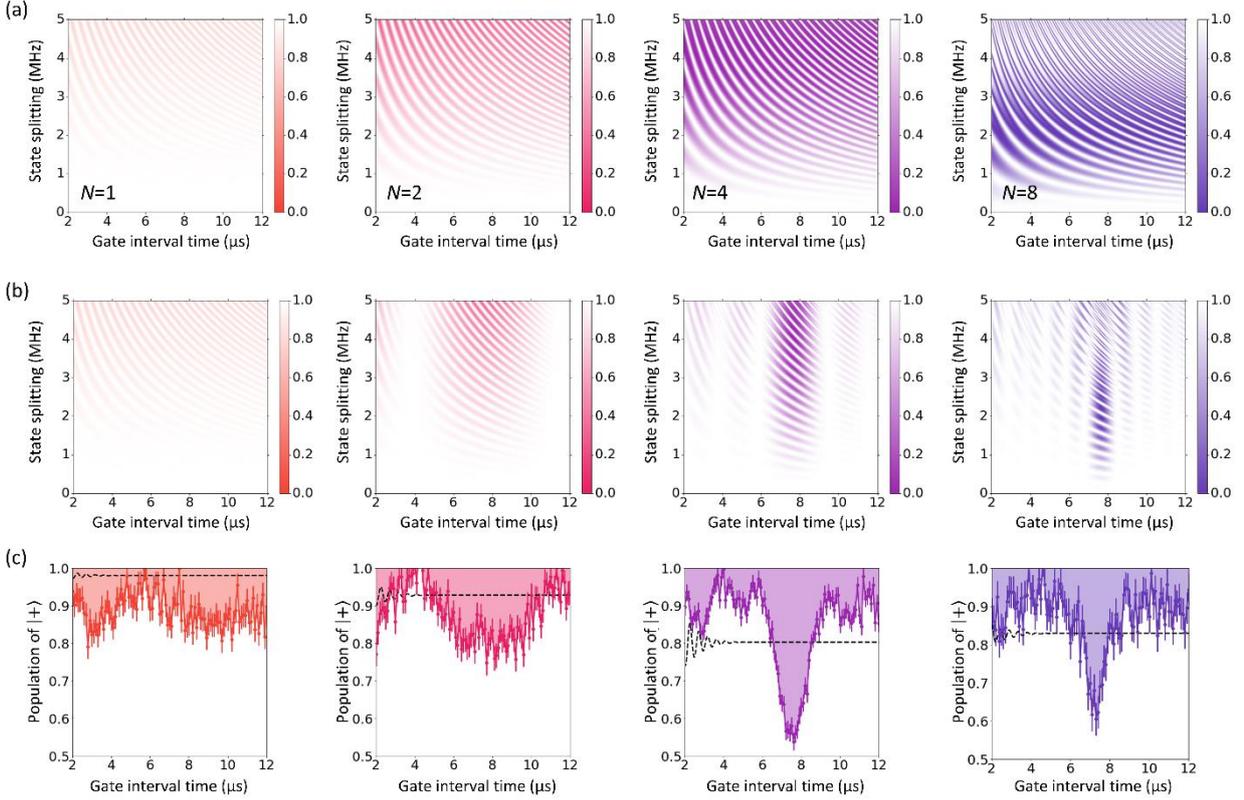

FIG. 2 (a, b) Numerical simulations for the dynamical decoupling of the geometric qubit with $N = 1$, 2, 4, and 8 gates depending on the state splitting between qubit basis states under (a) $\varDelta = 0$ and (b) $\varDelta = 130$ kHz. Color gradations represent the population of $|+\rangle$ after the decoupling. The state leakage that the state splitting induces is suppressed by introducing the detuning. (c) Experimental demonstration of dynamical decoupling of the geometric qubit with detuning. Solid lines show the experimental results and shaded regions show the leaked population of $|+\rangle$. Dotted lines show the numerical simulations under $\varDelta = 0$ as a comparison, assuming inhomogeneous broadening of state splitting caused by the carbon nuclear spin bath in a Gaussian distribution (the full width at 1/e is 0.3 MHz) as well as the hyperfine splitting caused by the nitrogen nuclear spin (2.2 MHz). Error bars indicate standard deviation.

The state fidelity of the geometric qubit is kept about the same even when the gate number of dynamical decoupling increases to 128, as shown in Fig. 3a. Fitting of the fidelity to $(1-\varepsilon_0)(1-\varepsilon_{\text{gate}})^N$ shows that the additional error per gate $\varepsilon_{\text{gate}}$ is as small as 0.03±0.03%, while the fundamental error including state preparation and measurement error $\varepsilon_0$ is as large as 10.2±0.6%. Here, the state fidelity is given by the average of $\text{Tr}(\rho_{\text{exp}}\rho_{\text{theo}})$ over the non-decohered area of the dynamical decoupling, where $\rho_{\text{exp}}$ and $\rho_{\text{theo}}$ denote experimental and theoretical density operators. Therefore, the coherence time can be appropriately estimated by the coherence envelope normalized by the fidelity; this envelope is expressed as $\exp\left(-\left(\frac{\tau}{T_2}\right)^p\right)$, where $p$ ranges from 1 to 4 (Fig. 3b). The obtained maximum coherence time is 1.9 ms, which is contributed by the spin relaxation time $T_1$. The geometric spin with the V-shaped system obeys the following relation, where the relaxation time $T_1$ (in this case, the transition caused by the leak of the green laser for spin initialization is also included) is involved between the qubit and the ancillary state instead of within the qubit space as for the typical qubit:

$$\frac{1}{T_2} = \frac{1}{T_2^{\text{pure}}} + \frac{1}{T_1}, \tag{1}$$

where $T_2^{\text{pure}}$ is the pure coherence time conditioned on the qubit space and $T_2$ is the experimentally observed coherence time. Note that the maximum value of $T_2$ is limited by $T_1$ instead of $2T_1$ for the conventional dynamic qubit. Figure 3c shows the convergence of $T_2$ to $T_1 \sim 2.6$ ms (Fig. 3c inset) and $T_2^{\text{pure}}$ derived from Eq. (1). We find that $T_2^{\text{pure}}$ increases linearly in proportion to the gate number by fitting the derived data with a power scaling of $T_2^{\text{pure}} \propto N^{1\pm0.09}$, which is consistent with the non-Markovian nature of the nuclear spin bath, where the inter-bath interaction is weak enough [30,31].

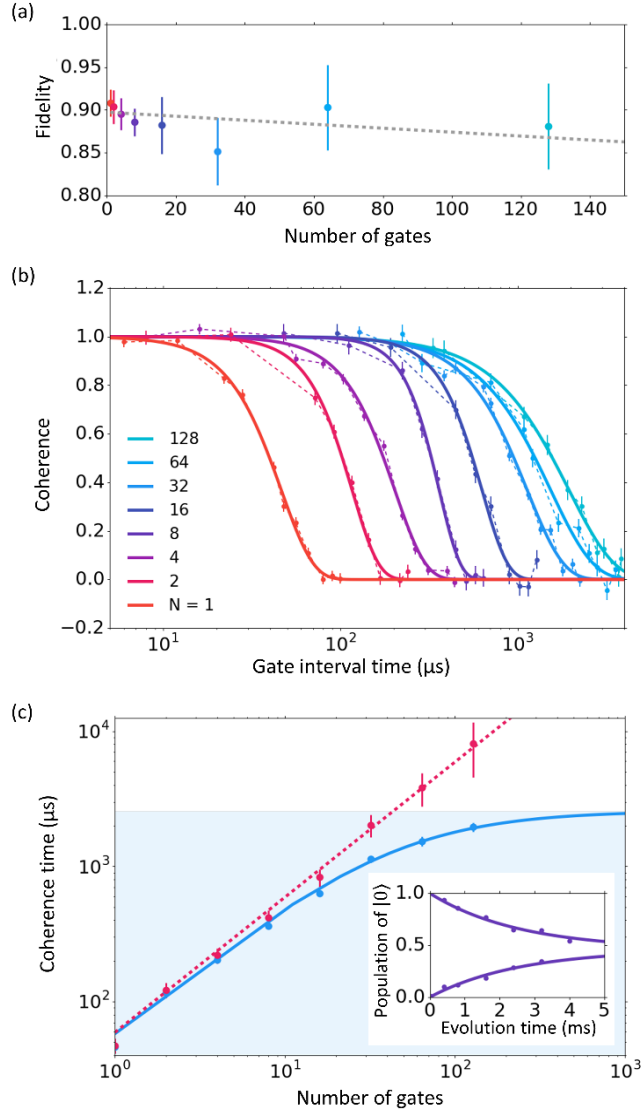

FIG. 3 (a) State fidelity after dynamical decoupling of the geometric qubit initially in $|+\rangle$ depending on the number of gates from 1 to 128. The dashed line shows exponential decay fitting. (b) The coherence envelope normalized by the state fidelity. Solid lines show the fitted decay curve with $\exp\left(-\left(\frac{\tau}{T_2}\right)^p\right)$. (c) The coherence time depending on the number of gates. $T_2$ derived from (b) (blue dots) is limited to $T_1$ (blue region), which is obtained from population decay between $|\pm\rangle$ and $|0\rangle$ (inset). $T_2^{\text{pure}}$ (red dots) is estimated by assuming Eq. (1) and linearly fitted (red dashed line). The linear fitting is reconverted to $T_2$ by Eq. (1) (blue solid line). Error bars indicate standard deviation.

The previously demonstrated robustness against electric field noise and temperature drift [32], and enhancement of magnetic field sensitivity [33] in double quantum qubit utilizing ms=±1 states in a three-level system can also be adapted to our qubit. Moreover, it has been demonstrated that the degeneracy of our qubit is required for interfacing with photon polarization [23,24], and that the coherence time in a low magnetic field regime is maximized under a zero magnetic field [9,31]. Although the xy-decoupling scheme shown in Ref. [33] also shows similar operational-error-tolerance, the developed scheme is easily implemented and adaptable to other three-level systems regardless of degeneracy. However, neither approach is completely independent of the prepared state. Thus, it may be necessary to introduce another rotation axis in a manner similar to the xy-4 sequence [2], in our case based on the polarization instead of the global phase of the microwave, in order to compensate for anisotropy. Pulse-shaping techniques, including the composite pulse, will be needed in order to further increase the gate fidelity.

Although the $T_1$ relaxation is inevitable, it can be suppressed remarkably by lowering the temperature. The relaxation-induced error can be eliminated in principle by heralding the relaxation event via the single-shot measurement of the population in the ancillary |0⟩ state without demolishing the qubit state. The heralding technique increases fidelity drastically not only for a measurement-based quantum computer but also for a quantum repeater, which requires probabilistic inter-node entanglement generation.

The extended coherence time contributes not only to the long lifetime of quantum memories used for quantum computers, quantum simulators, and quantum repeaters but also to the extreme sensitivity of quantum sensors against the DC electric field, AC magnetic field, and temperature under a zero magnetic field. The developed dynamical decoupling of the geometric qubit would also introduce a new approach for engineering a robust quantum gate sequence to implement large-scale quantum information processing.

**ACKNOWLEDGEMENTS**

We thank Yuichiro Matsuzaki, Burkhard Scharfenberger, Kae Nemoto, William Munro, Norikazu Mizuochi, Nobuyuki Yokoshi, Fedor Jelezko and Joerg Wrachtrup for their discussions and experimental help. This work was supported by Japan Society for the Promotion of Science (JSPS) Grants-in-Aid for Scientific Research (16H06326, 16H01052, 16K13818); by the Ministry of Education, Culture, Sports, Science, and Technology (MEXT) of Japan as an "Exploratory Challenge on Post-K computer" (Frontiers of Basic Science: Challenging the Limits); by the Research Foundation for Opto-Science and Technology; and by a Japan Science and Technology Agency (JST) CREST Grant (JPMJCR1773).

# References

[1] H. Y. Carr and E. M. Purcell, Effects of diffusion on free precession in nuclear magnetic resonance experiments, Phys. Rev. **94**, 630 (1954).

[2] T. Gullion, D. B. Baker, and M. S. Conradi, New, compensated Carr-Purcell sequences, J. Magn. Reson. **89**, 479 (1990).

[3] C. A. Ryan, J. S. Hodges, and D. G. Cory, Robust decoupling techniques to extend quantum coherence in diamond, Phys. Rev. Lett. **105**, 200402 (2010).

[4] A. Reiserer, N. Kalb, M. S. Blok, K. J. M. van Bemmelen, T. H. Taminiau, R. Hanson, D. J. Twitchen, and M. Markham, Robust quantum-network memory using decoherence-protected subspaces of nuclear spins, Phys. Rev. X **6**, 021040 (2016).

[5] G. de Lange, Z. H. Wang, D. Ristè, V. V Dobrovitski, and R. Hanson, Universal dynamical decoupling of a single solid-state spin from a spin bath, Science **330**, 60 (2010).

[6] D. Farfurnik, A. Jarmola, L. M. Pham, Z. H. Wang, V. V. Dobrovitski, R. L. Walsworth, D. Budker, and N. Bar-Gill, Optimizing a dynamical decoupling protocol for solid-state electronic spin ensembles in diamond, Phys. Rev. B **92**, 060301 (2015).

[7] N. Bar-Gill, L. M. Pham, A. Jarmola, D. Budker, and R. L. Walsworth, Solid-state electronic spin coherence time approaching one second, Nat. Commun. **4**, 1743 (2013).

[8] M. H. Abobeih, J. Cramer, M. A. Bakker, N. Kalb, M. Markham, D. J. Twitchen, and T. H. Taminiau, One-second coherence for a single electron spin coupled to a multi-qubit nuclear-spin environment, Nat. Commun. **9**, 2552 (2018).

[9] Y. Sekiguchi, Y. Komura, S. Mishima, T. Tanaka, N. Niikura, and H. Kosaka, Geometric spin echo under zero field, Nat. Commun. **7**, 11668 (2016).

[10] Y. Sekiguchi, N. Niikura, R. Kuroiwa, H. Kano, and H. Kosaka, Optical holonomic single quantum gates with a geometric spin under a zero field, Nat. Photonics **11**, 209 (2017).

[11] N. Ishida, T. Nakamura, T. Tanaka, S. Mishima, H. Kano, R. Kuroiwa, Y. Sekiguchi, and H. Kosaka, Universal holonomic single quantum gates over a geometric spin with phase-modulated polarized light, Opt. Lett. **43**, 2380 (2018).

[12] K. Nagata, K. Kuramitani, Y. Sekiguchi, and H. Kosaka, Universal holonomic quantum gates over geometric spin qubits with polarised microwaves, Nat. Commun. **9**, 3227 (2018).

[13] M. V Berry, Quantal phase factors accompanying adiabatic changes, Proc. R. Soc. A Math. Phys. Eng. Sci. **392**, 45 (1984).

[14] C. G. Yale, F. J. Heremans, B. B. Zhou, A. Auer, G. Burkard, and D. D. Awschalom, Optical manipulation of the Berry phase in a solid-state spin qubit, Nat. Photonics **10**, 184 (2016).

[15] B. B. Zhou, A. Baksic, H. Ribeiro, C. G. Yale, F. J. Heremans, P. C. Jerger, A. Auer, G. Burkard, A. A. Clerk, and D. D. Awschalom, Accelerated quantum control using superadiabatic dynamics in a solid-state lambda system, Nat. Phys. **13**, 330 (2017).

[16] E. Sjöqvist, D. M. Tong, L. Mauritz Andersson, B. Hessmo, M. Johansson, and K. Singh, Non-adiabatic holonomic quantum computation, New J. Phys. **14**, 103035 (2012).

[17] S. Arroyo-Camejo, A. Lazariev, S. W. Hell, and G. Balasubramanian, Room temperature high-fidelity holonomic single-qubit gate on a solid-state spin, Nat. Commun. **5**, 4870 (2014).

[18] C. Zu, W.-B. Wang, L. He, W.-G. Zhang, C.-Y. Dai, F. Wang, and L.-M. Duan, Experimental realization of universal geometric quantum gates with solid-state spins, Nature

(London) **514**, 72 (2014).

[19] B. B. Zhou, P. C. Jerger, V. O. Shkolnikov, F. J. Heremans, G. Burkard, and D. D. Awschalom, Holonomic Quantum Control by Coherent Optical Excitation in Diamond, Phys. Rev. Lett. **119**, 140503 (2017).

[20] Y. Wu, F. Jelezko, M. B. Plenio, and T. Weil, Diamond Quantum Devices in Biology, Angew. Chemie Int. Ed. **55**, 6586 (2016).

[21] F. Casola, T. Van Der Sar, and A. Yacoby, Probing condensed matter physics with magnetometry based on nitrogen-vacancy centres in diamond, Nat. Rev. Mater. **3**, 17088 (2018).

[22] E. Togan, Y. Chu, A. S. Trifonov, L. Jiang, J. Maze, L. Childress, M. V. G. Dutt, A. S. Sørensen, P. R. Hemmer, A. S. Zibrov, and M. D. Lukin, Quantum entanglement between an optical photon and a solid-state spin qubit, Nature (London) **466**, 730 (2010).

[23] H. Kosaka and N. Niikura, Entangled Absorption of a Single Photon with a Single Spin in Diamond, Phys. Rev. Lett. **114**, 053603 (2015).

[24] S. Yang, Y. Wang, D. D. B. Rao, T. Hien Tran, A. S. Momenzadeh, M. Markham, D. J. Twitchen, P. Wang, W. Yang, R. Stöhr, P. Neumann, H. Kosaka, and J. Wrachtrup, High-fidelity transfer and storage of photon states in a single nuclear spin, Nat. Photonics **10**, 507 (2016).

[25] B. Hensen, H. Bernien, A. E. Dréau, A. Reiserer, N. Kalb, M. S. Blok, J. Ruitenberg, R. F. L. Vermeulen, R. N. Schouten, C. Abellán, W. Amaya, V. Pruneri, M. W. Mitchell, M. Markham, D. J. Twitchen, D. Elkouss, S. Wehner, T. H. Taminiau, and R. Hanson, Loophole-free Bell inequality violation using electron spins separated by 1.3 kilometres, Nature (London) **526**, 682 (2015).

[26] G. Xu and G. Long, Universal nonadiabatic geometric gates in two-qubit decoherence-free subspaces, Sci. Rep. **4**, 6814 (2014).

[27] S. B. Zheng, C. P. Yang, and F. Nori, Comparison of the sensitivity to systematic errors

between nonadiabatic non-Abelian geometric gates and their dynamical counterparts, Phys. Rev. A **93**, 032313 (2016).

[28] See Supplemental Material at [URL will be inserted by publisher] for additional information on population leakage of geometric qubit.

[29] V. M. Acosta, E. Bauch, M. P. Ledbetter, A. Waxman, L.-S. Bouchard, and D. Budker, Temperature Dependence of the Nitrogen-Vacancy Magnetic Resonance in Diamond, Phys. Rev. Lett. **104**, 070801 (2010).

[30] M. J. Biercuk and H. Bluhm, Phenomenological study of decoherence in solid-state spin qubits due to nuclear spin diffusion, Phys. Rev. B **83**, 235316 (2011).

[31] N. Zhao, S.-W. Ho, and R.-B. Liu, Decoherence and dynamical decoupling control of nitrogen vacancy center electron spins in nuclear spin baths, Phys. Rev. B **85**, 115303 (2012).

[32] K. Fang, V. M. Acosta, C. Santori, Z. Huang, K. M. Itoh, H. Watanabe, S. Shikata, and R. G. Beausoleil, High-Sensitivity Magnetometry Based on Quantum Beats in Diamond Nitrogen-Vacancy Centers, Phys. Rev. Lett. **110**, 130802 (2013).

[33] H. J. Mamin, M. H. Sherwood, M. Kim, C. T. Rettner, K. Ohno, D. D. Awschalom, and D. Rugar, Multipulse Double-Quantum Magnetometry with Near-Surface Nitrogen-Vacancy Centers, Phys. Rev. Lett. **113**, 030803 (2014).
14